\documentclass[aps,prd,preprint,nofootinbib]{revtex4-2} 
\usepackage{graphicx,amsmath,amsfonts,amssymb,revsymb,dcolumn} 
\usepackage[utf8]{inputenc}
\input{epsf}
\usepackage{verbatim}

\newcommand{\be}{\begin{equation}}
\newcommand{\ee}{\end{equation}}
\newcommand{\bea}{\begin{eqnarray}}
\newcommand{\eea}{\end{eqnarray}}

\begin{document}


\title{Breakdown of the connection between symmetries and conservation laws for semiholonomic systems}
\author{Nivaldo A.  \surname{Lemos}}
\affiliation{Instituto de Física, Universidade Federal Fluminense,
Campus da Praia Vermelha, Niterói, 24210-340, RJ, Brazil.\footnote{
Electronic mail:  nivaldolemos@id.uff.br}}

\date{\today}

\nopagebreak


\begin{abstract}
\noindent Integrable velocity-dependent constraints are said to be semiholonomic. For good reasons, holonomic and semiholonomic  constraints are thought to be indistinguishable in Lagrangian mechanics. This well-founded belief notwithstanding, here we show by means of an example and a broad analysis that  the connection between symmetries and conservation laws, which holds for holonomic systems, is not valid  in general for systems subject to semiholonomic constraints.
\end{abstract}

\nopagebreak


\begin{abstract}
\noindent Integrable velocity-dependent constraints are said to be semiholonomic. For good reasons, holonomic and semiholonomic  constraints are thought to be indistinguishable in Lagrangian mechanics. This well-founded belief notwithstanding, here we show by means of an example and a broad analysis that  the connection between symmetries and conservation laws, which holds for holonomic systems, is not always valid   for systems subject to semiholonomic constraints.
\end{abstract} 

\maketitle

\section{Introduction}

The connection between symmetries and conservation laws is fundamental not only  in classical mechanics \cite{Goldstein} but also in quantum mechanics \cite {Schiff} and quantum field theory \cite{Bogoliubov}. In classical mechanics, Noether's theorem \cite{Noether} establishes the most general correspondence between invariance under continuous transformations and constants of the motion. The conservation of linear momentum, angular momentum and energy for many-particle systems is associated with invariance of the action under translations, rotations and time displacements, respectively \cite{Goldstein,Desloge,Hanc}. This connection generalizes to {\it holonomically} constrained systems as long as the Lagrangian {\it and} the  constraints are invariant under the mentioned transformations \cite{Lemos}. Integrable velocity-dependent constraints seem, for all intents and purposes,  to be equivalent to holonomic constraints because, in their integrated form, they impose restrictions on the coordinates alone. The goal of this note is to show, by means of an example and a general analysis, that integrable velocity-dependent constraints are not equivalent to holonomic constraints as far as the connection between symmetries and conservation laws is concerned.

\section{An Unexpected Example}

Nonintegrable velocity-dependent constraints are said to be nonholonomic, whereas integrable  velocity-dependent constraints are called semiholonomic \cite{Flannery,Papastavridis}. A mechanical system is  semiholonomic if  it is subject to  velocity-dependent  constraints all of which are integrable either separately  or  when taken together.  The mechanical system we study from now on is semiholonomic.

\begin{figure}[t!]
\epsfxsize=8cm
\begin{center}
\leavevmode
\epsffile{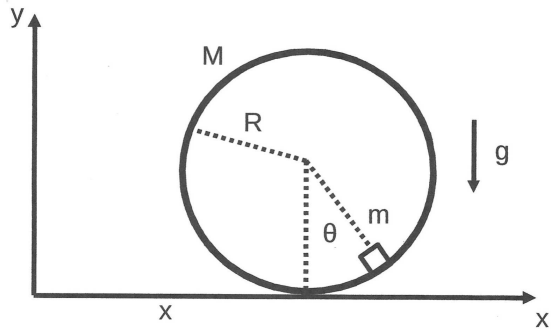}
\end{center}
\caption{The  block slides without friction on the inner surface of the cylindrical shell that rolls without slipping on a horizontal plane.}
\label{Figure1}
\end{figure}

Consider a small block of mass $m$ that slides without friction on the inner surface of a cylindrical shell of mass $M$ and radius $R$, as in  Fig. \ref{Figure1}. The system is in a uniform gravitational field and the cylinder  rolls without slipping on a horizontal surface.

Let us construct a Lagrangian for this system with the use of the coordinates $x$ and $\theta$ shown in Fig. \ref{Figure1}, where $x$ gives the position of the cylinder's center of mass with respect to the origin of the chosen inertial reference frame. The kinetic energy of the cylinder is
\begin{equation}
\label{kinetic-cyl}
T_{cyl} = \frac{M}{2}{\dot x}^2 + \frac{I_{CM}}{2}\omega^2 = M {\dot x}^2
\end{equation}
where we have used $I_{CM}=MR^2$ and the rolling constraint ${\dot x} = R \omega $, where $\omega $ is the cylinder's angular velocity about its symmetry axis. Letting $X,Y$ be the Cartesian coordinates of the block, we have
\begin{equation}
\label{cartesian-block}
X  = x+ R \sin \theta , \qquad Y = R - R \cos \theta ,
\end{equation}
whence
\begin{equation}
\label{cartesian-veloc-block}
{\dot X}  = {\dot x} + R {\dot \theta} \cos \theta , \qquad {\dot Y} = R {\dot \theta} \sin \theta .
\end{equation}
Thus, the block's kinetic energy is
\begin{equation}
\label{kinetic-block}
T_{block} = \frac{m}{2}({\dot X}^2 +  {\dot Y}^2) = \frac{m}{2}{\dot x}^2 + \frac{m R^2}{2} {\dot \theta}^2 + mR{\dot x}{\dot \theta}\cos \theta . 
\end{equation}
The only varying  potential energy is the block's, given by $V= mgY = mgR - mgR\cos \theta$. Since $L= T_{cyl}+T_{block} -V$, it follows that the  Lagrangian is 
\begin{equation}
\label{Lagrangian}
L = \frac{2M+m}{2}{\dot x}^2 + \frac{mR^2}{2} {\dot \theta}^2 + mR{\dot x}{\dot \theta} \cos \theta + mgR\cos \theta ,
\end{equation}
where an unimportant additive constant has been dropped. 

Lagrange's equations read
\begin{eqnarray}
\label{Lagrange-eq1}
(2M+m){\ddot x} + mR {\ddot \theta} \cos \theta - mR {\dot \theta}^2 \sin \theta & = & 0 , \\
\label{Lagrange-eq2}
mR^2 {\ddot \theta} + mR {\ddot x} \cos \theta + mgR \sin \theta & = & 0.
\end{eqnarray}
Since $x$ is a cyclic coordinate of the Lagrangian \eqref{Lagrangian}, its conjugate momentum $p_x = \partial L/\partial {\dot x}$ is a constant of the motion:
\begin{equation}
\label{px-constant}
(2M+m){\dot x} + mR {\dot \theta} \cos \theta = p_x = \mbox{constant} . 
\end{equation}
The equation  of motion \eqref{Lagrange-eq1} simply expresses the fact that $dp_x/dt =0$. Besides $p_x$,  the total energy, $E$,  is a second independent constant of the motion. Therefore, this two-degree-of-freedom system is integrable --- see Section 9.7 of  \cite{Lemos}. The solution to the equations of motion by quadratures will most likely involve elliptic or hyperelliptic functions. Finding such a solution is not the main purpose of this note, so we will not pursue this here.

The Lagrangian \eqref{Lagrangian} does not depend on the cylinder's center of mass coordinate $x$, therefore it is invariant under translations along the $x$-axis.   In view of the connection between translational invariance along a certain direction and conservation of the corresponding component of the total linear momentum, it would be natural to expect  the constant of the motion \eqref{px-constant} to coincide with $P_x$, the $x$-component of the total linear momentum. But, upon using \eqref{cartesian-veloc-block}, we find that
\begin{equation}
\label{Px}
P_x = M{\dot x} + m{\dot X} = (M+m){\dot x} + mR {\dot \theta} \cos \theta ,
\end{equation}
 which is not equal to the canonical momentum  $p_x$ and {\it is not a constant of the motion}. Therefore, in the present case, invariance under translations does not imply conservation of the total linear momentum.

On second thought, the nonconservation of $P_x$ is reasonable because, in order for the cylinder to roll without slipping,  there must be friction between the cylinder and the horizontal surface. Since, from Newton's second and third  laws, $dP_x/dt = f_s$ where $f_s$ is the static frictional force exerted by the horizontal surface on the cylinder, which is the total horizontal external force on the system, $P_x$ cannot remain constant.

A general theorem  \cite{Lemos} states that if the Lagrangian takes the standard form $L=T-V$, where $V$ does not depend on the velocities, and both the Lagrangian and the constraints, all of them holonomic, are invariant under translations, then the total linear momentum is conserved. The reason for the failure of this result in the present case appears to be the existence of the velocity-dependent constraint ${\dot x} - R \omega =0$, even though it is integrable. Letting $\phi$ be the rotation angle (positive clockwise) of the cylinder about its symmetry axis, then $\omega = \dot \phi$ and the rolling constraint immediately integrates to $x - R \phi - C = 0$, with $C$ an arbitrary constant. In the case of an infinitesimal translation along the $x$-axis, $\delta x \neq 0$ and $\delta \phi =0$. Therefore, in its integrated form, the rolling constraint
is not invariant under translations because
\begin{equation}
\label{constraint-not-invariant}
\delta  (x - R \phi - C) = \delta x - R \delta \phi - \delta C= \delta x \neq 0.
\end{equation}

Another way to see this invariance failure is by noting that an extended Lagrangian that  yields the correct equations of motion and the rolling constraint at a single stroke is given by \cite{Lemos}
\begin{equation}
\label{extended-Lagrangian}
{\cal L} = {\tilde L} + \lambda (x - R \phi - C),
\end{equation}
where the Lagrange multiplier $\lambda$ is treated  as an additional coordinate and $\tilde L$ is the unconstrained Lagrangian 
\begin{equation}
\label{unconstrained-Lagrangian}
{\tilde L} = \frac{M+m}{2}{\dot x}^2 + \frac{MR^2}{2}{\dot \phi}^2 + \frac{mR^2}{2} {\dot \theta}^2 + mR{\dot x}{\dot \theta} \cos \theta + mgR\cos \theta ,
\end{equation}
which is set up without using the rolling constraint ${\dot x}=R{\dot \phi}$. Lagrange's equations
\begin{equation}
\label{extended-Lagrange-equations}
\frac{d}{dt}\bigg( \frac{\partial {\cal L}}{\partial {\dot \xi}_k} \bigg) -  \frac{\partial {\cal L}}{\partial \xi_k} =0
\end{equation}
for $\xi_1=x, \xi_2=\phi ,\xi_3 = \theta , \xi_4= \lambda$ are
\begin{eqnarray}
\label{extended-Lagrange-equations-explicit}
\label{extended-eq1}
(M+m) {\ddot x} + mR{\ddot \theta} \cos \theta - mR {\dot \theta}^2 \sin \theta & = & \lambda , \\
\label{extended-eq2}
MR^2{\ddot \phi} & = & - R \lambda , \\
\label{extended-eq3}
mR^2{\ddot \theta} + mR {\ddot x} \cos \theta + mgR \sin \theta & = & 0 , \\
\label{extended-eq4}
x-R\phi - C & = & 0 .
\end{eqnarray}
Equation \eqref{extended-eq3} is the same as equation \eqref{Lagrange-eq2}. With the use of  \eqref{extended-eq4} equation \eqref{extended-eq2}
becomes $M{\ddot x}= -\lambda$. Once this is inserted into \eqref{extended-eq1} one reproduces equation \eqref{Lagrange-eq1}. Thus, the extended Lagrangian  $\cal L$  yields the correct  equations of motion. Equations \eqref{extended-Lagrange-equations} are sometimes referred to as the  d’Alembert generalized principle \cite{Flannery}.

It is clear that the extended Lagrangian $\cal L$ defined by \eqref{extended-Lagrangian} is not invariant under translations along the $x$-axis because, although the unconstrained Lagrangian $\tilde L$ does not depend on $x$, the added term containing the Lagrange multiplier does. As a consequence, the $x$-component of the total linear momentum is not conserved. It should be noted that for the unconstrained Lagrangian \eqref{unconstrained-Lagrangian} the canonical momentum conjugate to $x$  is equal to $P_x$, the $x$-component of the total linear momentum, as expected. However, the use of the rolling constraint ${\dot x} - R{\dot \phi}=0$ to go from the unconstrained Lagrangian $\tilde L$ to the reduced Lagrangian $L$ given by Eq. \eqref{Lagrangian} adds a contribution to  the  momentum conjugate to $x$,  making $p_x$  different from $P_x$. 

By the way, the static friction force required to enforce the rolling constraint is $f_s =  \lambda$ inasmuch as the left-hand side of \eqref{extended-eq1} is equal to $dP_x/dt$, as follows from \eqref{Px}. As is the case for all forces in Newtonian mechanics, the constraint force  must be expressed exclusively in terms of coordinates, velocities and time {\cite{Moriconi}. Solving equations \eqref{Lagrange-eq1} and \eqref{Lagrange-eq2} for $\ddot x$ and $\ddot \theta$ and inserting their expressions in terms of $\theta$ and $\dot \theta$       into  \eqref{extended-eq1}, $f_s$ is properly determined, if one so wishes.

An alternative way to treat the   problem is by introducing the new extended Lagrangian ${\tilde {\cal L}} = {\tilde L} + \Lambda ({\dot x} - R \dot \phi )$,  which accounts for the constraint in its velocity-dependent form. It can be easily checked that $\tilde {\cal L}$ gives rise to the same equations of motion as the above $\cal L$ with the new Lagrange multiplier $\Lambda$ related to the old one by ${\dot \Lambda}= - \lambda$. Now ${\tilde {\cal L}}$ does not depend either on $x$ or $\phi$, implying that  ${\tilde p}_x = \partial {\tilde {\cal L}}/\partial {\dot x} = (M+m) {\dot  x} + mR {\dot \theta} \cos \theta + \Lambda$ and ${\tilde p}_{\phi} = \partial {\tilde {\cal L}}/\partial  {\dot \phi} = MR^2 {\dot \phi} - \Lambda R$ are constants of the motion. These two constants of the motion depend on the Lagrange multiplier, but with the use of the constraint ${\dot x} - R{\dot \phi}=0$ one finds that the combination ${\tilde p}_x + {\tilde p}_{\phi}/R$ yields the constant of the motion \eqref{px-constant}. This sheds light on the meaning of the conserved canonical momentum $p_x$, which appears as a mixture of linear and angular momenta induced  by the constraint.  This distinct formulation is advantageous if the constraints are known to be integrable but their expression in integrated form is hard to find. This may happen in the event that there are two or more velocity-dependent constraints which are integrable when taken together but are  not separately integrable.


In order to broaden the scope of our discussion, we proceed to a general analysis of semiholonomic systems that may  behave like  the mechanical system that has just been studied.

\section{A General Analysis}

Let a mechanical system be described by the $N$ coordinates $q_1, \ldots , q_N$ and  the unconstrained Lagrangian
\begin{equation}
\label{general-unconstrained-L}
{\tilde L} = {\tilde L}(q_1, \ldots , q_N, {\dot q}_1, \ldots , {\dot q}_N, t),
\end{equation} 
which is set up as if there were no constraints. Suppose the system is subject to the integrable velocity-dependent constraints
\begin{equation}
\label{general-velocity-constraints}
g_l  (q_1, \ldots , q_N, {\dot q}_1, \ldots , {\dot q}_N, t) = 0,\qquad l=1, \ldots , p
\end{equation} 
where $p<N$ and the functions $g_l$ are assumed to depend linearly on the velocities. The constraints are integrable (see Appendix B of {\cite{Lemos}) if there exist a  velocity-independent everywhere nonsingular $p \times p$ matrix $(h_{lk})$ and velocity-independent functions $f_1, \ldots ,f_p$ such that
\begin{equation}
\label{integrable-velocity-constraints}
g_l  (q_1, \ldots , q_N, {\dot q}_1, \ldots , {\dot q}_N, t) = \sum_{k=1}^p h_{lk} (q_1, \ldots , q_N,t)\frac{d}{dt}f_k(q_1, \ldots , q_N,t), \quad l=1, \ldots , p.
\end{equation}
 One has 
\begin{equation}
\label{equivalence-g-f}
g_1= 0, \ldots ,  g_p =0 \qquad \Longleftrightarrow \qquad {\dot f}_1 =0,  \ldots , {\dot f}_p =0 
\end{equation}
because the matrix $(h_{lk})$ is invertible. Therefore the constraints \eqref{general-velocity-constraints} can be replaced by 
\begin{equation}
\label{integrated-velocity-constraints}
f_1 (q_1, \ldots , q_N,t) - C_1 =0, \ldots,   f_p (q_1, \ldots , q_N,t) - C_p =0
\end{equation} 
where $C_1, \ldots , C_p$ are arbitrary constants.
Setting $n=N-p$, let us  assume without loss of generality that the $p$ constraint equations \eqref{integrated-velocity-constraints} can be solved for $q_{N-p+1}, \ldots , q_{N}$ in terms of the remaining  variables $q_1, \ldots , q_n$. 

Because the constraints are integrable, upon their insertion into the unconstrained Lagrangian one obtains the reduced Lagrangian
\begin{equation}
\label{reduced-L}
L (q_1, \ldots , q_n, {\dot q}_1,  \ldots , {\dot q}_n, t) = {\tilde L}(q_1, \ldots , q_N, {\dot q}_1, \ldots , {\dot q}_N, t)\bigg\vert_{ g_1= \ldots = g_p =0}\, .
\end{equation}
The number of degrees of freedom is $n$ and  the coordinates $q_1, \ldots , q_n$ are mutually independent, that is, they are genuine generalized coordinates. The correct equations of motion for the semiholonomic system are the usual Lagrange equations
\begin{equation}
\label{usual-Lagrange-equations}
\frac{d}{dt}\bigg( \frac{\partial L}{\partial {\dot q}_i} \bigg) - \frac{\partial L}{\partial  q_i} =0, \qquad i=1, \ldots , n.
\end{equation}

These equations of motion are also generated by the extended Lagrangian 
\begin{equation}
\label{extended-L}
{\cal L} (\xi, {\dot \xi} , t) = {\tilde L}(q_1, \ldots , q_N, {\dot q}_1, \ldots , {\dot q}_N, t) + \sum_{l=1}^p \lambda_l \bigl( f_l (q_1, \ldots , q_N,t) - C_l\bigr)
\end{equation}
in which $\xi = (q_1, \ldots , q_N, \lambda_1, \ldots , \lambda_p)$ is considered as a set of $N+p$ independent variables. The equations of motion
\begin{equation}
\label{extended-Lagrange-equations-general}
\frac{d}{dt}\bigg( \frac{\partial \cal L}{\partial {\dot \xi}_s} \bigg) - \frac{\partial \cal L}{\partial  \xi_s} =0, \qquad s=1, \ldots , N+p 
\end{equation}
yield
\begin{equation}
\label{Lagrange-equation-constrained}
\frac{d}{dt}\bigg( \frac{\partial {\tilde L}}{\partial {\dot q}_k} \bigg) - \frac{\partial {\tilde L}}{\partial  q_k} =\sum_{l=1}^p \lambda_l \frac{\partial f_l}{\partial q_k}, \qquad k=1, \ldots , N
\end{equation}
as well as the constraint equations \eqref{integrated-velocity-constraints}. Once  the constraints  \eqref{integrated-velocity-constraints} are used, equations \eqref{Lagrange-equation-constrained} coincide with the reduced Lagrange equations \eqref{usual-Lagrange-equations}. 

Now, for the completion of our analysis: Let $q_j$, for some fixed index $j \in \{1, \ldots , n\}$, be a cyclic coordinate of the unconstrained Lagrangian $\tilde L$ and of the velocity-dependent constraints \eqref{general-velocity-constraints}. This means that $\tilde L$ and all of the  constraint functions $g_l$ do not depend on $q_j$, but both $\tilde L$ and at least one of the constraint functions $g_l$ depend on ${\dot q}_j$. Since neither  $\tilde L$ nor the constraint functions $g_l$ depend on $q_j$, the reduced Lagrangian $L$ given by equation \eqref{reduced-L} does not depend on $q_j$. Therefore, the reduced Lagrangian is invariant under displacements of $q_j$ and the canonical momentum $p_j= \partial L/\partial {\dot q}_j$ is a constant of the motion.
On the other hand, because of \eqref{integrable-velocity-constraints}, at least one of the $f_l$ depends on $q_j$ since at least one of the $g_l$ depends on  ${\dot q}_j$. Therefore, the extended Lagrangian \eqref{extended-L} depends on $q_j$ and it  follows that $P_j=\partial {\cal L}/\partial {\dot q}_j$ is not a constant of the motion. This occurs because $q_j$ does not appear in the constraints  in their  velocity-dependent form but does appear in their integrated form. This is exactly what happened in our example. In its velocity-dependent form the constraint ${\dot x}-R{\dot \phi}=0$ does not depend on $x$, but in the integrated form $x-R\phi -C=0$ it does. 
As it turned out,  the conserved canonical momentum $p_x$ is different from $P_x$, the $x$-component of the total linear momentum,  which is not conserved.

As a general rule, since $P_j=\partial {\cal L}/\partial {\dot q}_j = \partial {\tilde L}/\partial {\dot q}_j$ where $\tilde L$ is the unconstrained Lagrangian, the nonconserved momentum $P_j$ is a  linear momentum component  if $q_j$ is a translational  coordinate or an  angular momentum component if $q_j$ is a rotational  coordinate.

\section{Conclusion}

Semiholonomic and holonomic constraints are indistinguishable in just about every way. In particular, both holonomic and semiholonomic constraints actually reduce the number of degrees of freedom of a mechanical system \cite{Papastavridis,Neimark}. This means that
the reduced Lagrangian, which is obtained by inserting the constraints into the unconstrained Lagrangian, gives rise to the correct equations of motion. Furthermore, just like holonomic constraints, semiholonomic constraints can  be correctly
treated by the d’Alembert generalized principle \cite{Flannery}. Nevertheless, here we have shown that, as far as the connection between symmetries and conservation laws is concerned, holonomic and semiholonomic constraints behave differently. If the Lagrangian $L = T -V$ and the semiholonomic  constraints {\it in their velocity-dependent form} are invariant under translations along a certain direction {\it it does not follow} that the corresponding component of the total linear momentum is a constant of the motion. The same applies to semiholonomic constraints that are invariant under rotations in their velocity-dependent form. One cannot, in general, extend the connection between symmetries and conservation laws, which holds for holonomic systems, to systems subject to semiholonomic constraints.

  \begin{acknowledgments}
The author is thankful to the anonymous reviewers and the editor Todd Springer for their careful reading and germane criticism, which helped to improve the paper.
\end{acknowledgments}


\begin{thebibliography}{99}

\bibitem{Goldstein}  H. Goldstein, C. P. Poole, Jr. and J. L. Safko, {\it Classical Mechanics}, 3rd ed. (Addison Wesley, San Francisco, 2001), chap. 2.

\bibitem{Schiff}  L. I.  Schiff, {\it Quantum Mechanics}, 3rd ed. (McGraw-Hill, New York, 1968), chap. 7.

\bibitem{Bogoliubov}  N. N. Bogoliubov and D. V. Shirkov, {\it Introduction to the Theory of Quantized Fields}, 3rd ed. (Wiley, New York, 1980), chap. 1.

\bibitem{Noether} E. Noether, {\it Invariante Variationsprobleme},   Nachrichten von der Gesellschaft der Wissenschaften zu
G\"ottingen, Mathematisch-Physikalische Klasse, {\bf 2}, 235-57 (1918). English translation by M. A. Tavel: arXiv:physics/0503066.

\bibitem{Desloge} E. A. Desloge and R. I. Karch, {\it Noether’s theorem in classical mechanics}, Am. J. Phys. {\bf 45}, 336-339 (1977).

\bibitem{Hanc} J. Hanc, S. Tuleja and M. Hancova, {\it Symmetries and conservation laws: Consequences of Noether’s theorem}, Am. J. Phys. {\bf 72}, 428-436 (2004).

\bibitem{Lemos}  N. A. Lemos, {\it Analytical Mechanics} (Cambridge University Press, Cambridge, 2018).

\bibitem{Flannery} M. R. Flannery, {\it The enigma of nonholonomic constraints}, Am. J.  Phys. {\bf 73}, 265-272 (2005).

\bibitem{Papastavridis}  J. G. Papastavridis, {\it Analytical Mechanics: A Comprehensive Treatise on the Dynamics of Constrained Systems} (World Scientific, Singapore, 2014), chap. 2.

\bibitem{Moriconi} N. A. Lemos and M.  Moriconi, {\it On the consistency of the Lagrange multiplier method in classical mechanics}, Am. J.  Phys. {\bf 89}, 776-782 (2021).

\bibitem{Neimark} Ju. I. Ne\u{\i}mark and N. A. Fufaev, {\it Dynamics of Nonholonomic Systems} (American Mathematical Society, Providence, RI,  1972), chap. 1.






\end{thebibliography}
\end{document}